\documentclass[preprint,12pt]{elsarticle}

\usepackage{epsfig}

\usepackage{amssymb}

\journal{Solid State Communications}

\begin{document}

\begin{frontmatter}

\title{Formation of transition metal hydrides at high pressures}


\author{Olga Degtyareva$^{1,*}$, John E. Proctor$^{1}$, Christophe
Guillaume$^{1}$, Eugene Gregoryanz$^{1}$ and Michael Hanfland$^{2}$}

\ead{o.degtyareva@ed.ac.uk; phone: +44(0)1248812982} \address{$^1$School
of Physics and Centre for Science at Extreme Conditions, University of
Edinburgh, Edinburgh EH9 3JZ, UK \\ $^{2}$ESRF, BP 220, Grenoble, France.}

\begin{abstract}

Silane (SiH$_{4}$) is found to (partially) decompose at pressures
above 50 GPa at room temperature into pure Si and H$_{2}$. The released
hydrogen reacts with surrounding metals in the diamond anvil cell to
form metal hydrides. A formation of rhenium hydride is observed after the
decomposition of silane. From the data of a previous experimental report
(Eremets \textit{et al.}, Science {\bf 319}, 1506 (2008)), the claimed
high-pressure metallic and superconducting phase of silane is identified
as platinum hydride, that forms after the decomposition of silane. These
observations show the importance of taking into account possible chemical
reactions that are often neglected in high-pressure experiments.

\end{abstract}

\begin{keyword}High pressure \sep synchrotron x-ray diffraction \sep
synthesis \sep metals

\PACS 61.50.Ks \sep 62.50.+p



code \sep code (2000 is the default)

\end{keyword}

\end{frontmatter}


\section{Introduction}

Pressure causes extraordinary changes in materials, reducing interatomic
distances and modifying their properties. This often provides a path
for synthesis of novel materials. Substantial progress has been made
in the high pressure synthesis of hydrides, carbides, borides, and
nitrides in recent years \cite{Antonov, Brazhkin, carbide, nitrides,
nitridesreview}, mainly due to the development of high-pressure
high-temperature experimental techniques. New materials can often be
obtained following a decomposition of the starting materials, as for
example in the high pressure synthesis of transition metal hydrides
that do not form at normal conditions. In this case, hydrogen-containing
compounds such as AlH$_{3}$, LiAlH$_{4}$ etc that decompose on thermal
compression releasing hydrogen are routinely used as internal hydrogen
sources \cite{Antonov, Fukai}. Decomposition and chemical reactions at
high pressure can be harnessed to develop new materials for technological
applications \cite{mcmillan}. These phenomena can be used to advantage
in many different respects, as for example was demonstrated for pure
hydrogen \cite{goncharovhydrogen} that was obtained using decomposition
and chemical reactions of other materials to study fundamental physical
properties of its dense state. On the other hand, the systematic studies
of binary phase diagrams showed that in these simple
systems the phase equilibrium changes under pressure allowing for
decomposition of existing phases and synthesis of new phases \cite{Ponyatovskii}. The negligence of this and the possible high
pressure chemical reactions in a studied sample can lead to erroneous
interpretation of results.

The hydrogen-dominated group IVa alloys (CH$_{4}$, SiH$_{4}$, GeH$_{4}$
etc.) have been a subject of much scientific interest as candidates
for high temperature superconductors in their dense metallic states. As
proposed by theoretical studies \cite{Ashcroft, Feng}, due to the chemical
pre-compression of hydrogen these materials may require pressures far
less than expected for pure hydrogen to enter metallic states. Recent
experiments reported metallization of SiH$_{4}$ (silane) above 50 GPa
\cite{Chen, Eremets} with a formation of a hexagonal close-packed (hcp)
structure of space group P6$_{3}$ (Ref. \cite {Eremets}). Above 100 GPa,
this metallic phase was claimed to transform to a transparent insulating
phase \cite{Eremets} with a volume increase of $\sim$25\% upon transition,
co-existing with the metallic phase up to at least 192 GPa. The observed
transparent insulating phase of silane \cite{Eremets} was identical to
the I4$_{1}$/a phase earlier predicted by ab-initio calculations to
be stable between 50 and 250 GPa, above which a metallic phase with
space group C2/c was predicted to become stable \cite{Pickard}. Here
we note that the positive volume change on pressure increase and the
co-existence of two thermodynamically stable phases with different volumes
at constant temperature and composition stoichiometry contradicts Le
Chatelier's principle \cite{Stanley}. Indeed, the recent \textit{ab
initio} calculations on the experimentally reported metallic P6$_{3}$
phase of silane showed that this structure is mechanically highly
unstable and is unlikely to form at these pressure conditions \cite{Kim,
Chen2, MartinezCanales}, suggesting that the P6$_{3}$ phase might be of
a different chemical composition \cite{MartinezCanales}. The analysis
of these arguments suggests that a chemical reaction needs to be taken
into account to explain the presence of the metallic hcp phase in silane.

Here we report a (partial) decomposition of silane at high pressures and room
temperature into pure Si and hydrogen, where released hydrogen reacts with
the surrounding metals in the diamond anvil cell chamber forming metal
hydrides. We find a formation of Re hydride after decomposition of silane
and reaction of hydrogen with the Re gasket, and identify the recently
reported metallic phase of silane \cite{Eremets} as PtH \cite{Hirao}.

\section{Experimental}

\indent Electronic grade 99.998+\% silane (SiH$_{4}$) from Sigma Aldrich
was used for cryogenic loading of diamond anvil cells in a dry argon
purged glove box by pre-cooling the cells in a liquid nitrogen bath. The
loading procedure was similar to previous experimental studies of
SiH$_{4}$ (Ref. \cite{Degtyareva}). Diamond anvil cells with 250$\mu$m and
100$\mu$m culets were used, with rhenium gaskets. Pressure was estimated
from the first-order Raman spectra of the diamond anvil \cite{Hanfland}
for some experiments, while other loadings contained a ruby chip for
measuring the pressure from the shift of the R$_{1}$ ruby fluorescence
line using the ruby scale from \cite{ruby}. Powder x-ray diffraction data
were mostly collected at the beamline ID09 at the European Synchrotron
Radiation Facility (ESRF), Grenoble, while the spectrum at 108 GPa,
the maximum pressure reached in this study, was collected at the
beamline ID27 of ESRF. A focused monochromatic beam was used, with
wavelengths 0.4130 \AA\/ and 0.3738 \AA, and the data were recorded on
a MAR image plate. Diffraction data were integrated azimuthally using
FIT2D \cite{fit2d}.  Five independent loadings of silane were used to
study the sample in the pressure range from 10 to 108 GPa with three
experiments performed at room temperature and two experiments made at
low temperature of 100 K using the on-line cryostat of the beamline ID09.

\section{Results and discussion}

Firstly we note that the x-ray diffraction pattern of the "metallic
hcp phase of silane" reported in Ref. \cite{Eremets} is identical to
the hcp phase of platinum hydride \cite{Hirao} (Fig. 1 a,b). Platinum
and hydrogen do not form compounds at normal conditions, however
PtH with a hcp structure can be synthesized under pressure at above
27 GPa as recently reported by Hirao \textit{et al.} \cite{Hirao},
stable to at least 42 GPa. The "metallic hcp phase of silane" has the
same crystallographic characteristics as that of PtH. The axial ratio
of the hcp silane is c/a=1.681 \cite{Eremets}, very close to that of PtH
(c/a=1.702) \cite{Hirao} and the atomic volume data of Ref. \cite{Eremets}
lie very close to that of PtH \cite{Hirao} (Fig. 2). The peculiarly
small compressibility of the proposed metallic hcp silane \cite{Eremets}
(Fig. 2) can be understood as the hcp phase of PtH that has similar
compressibility to platinum \cite{Hirao} (bulk modulus 274 GPa
\cite{Dewaele}). These characteristics identify the proposed hcp phase
of silane uniquely as PtH \cite{note}. Since the experimental set-up of Eremets
\textit{et al.} \cite{Eremets} included platinum wires and foil for
measuring conductivity, it is very probable that at around 50 GPa due to
decomposition and chemical reaction the silane sample released hydrogen
to form PtH.

The composition of the Pt hydride synthesised by Hirao \textit{et al.}
is proposed to be 50/50 \cite{Hirao}. Indeed, the atomic volume of
Pt at 42.9 GPa is 13.42 \AA$^{3}$ (Ref. \cite{Dewaele}) compared to
the atomic volume of Pt-H by Hirao \textit{et al.} of 15.77 \AA$^{3}$
(Ref. \cite{Hirao}). This gives a difference in atomic volume of metal and
its hydride of about 2.35 \AA$^{3}$. It is known that during the formation
of a hydride, the volume of a transition metal increases by $2.1\pm0.2$
\AA$^{3}$ for one hydrogen atom occupying interstitial octahedral sites
in the structure of transition metals \cite{Fukaibook} (the values up to
2.8 \AA$^{3}$ can be found in literature \cite{Baranovski}). This yields an
estimate of the composition of platinum hydride as PtH. The volume data
for the "metallic hcp phase of silane" \cite{Eremets} for the highest
pressure give approximately 2.1 \AA$^{3}$ for the difference between the
volumes of platinum and platinum hydride yielding an estimated composition
close to PtH. At lower pressures, a smaller difference ($<$2.0 \AA$^{3}$)
is observed between Pt and platinum hydride (Fig. 2), which indicates
that an unsaturated PtH$_{x}$ with $x<1$ is probably formed and that
the hydrogen content of PtH$_{x}$ might be increasing with pressure.

The question remains about superconductivity, as the "metallic
hcp phase of silane" is reported to be superconducting with a
strong pressure dependence of T$_c$, that reaches a maximum of 17 K
(see Fig. 2 in Ref. \cite{Eremets}). The superconducting state is
characterized by exactly zero electrical resistance (if measured by
a 4-probe technique) and the exclusion of the interior magnetic field
(the Meissner effect). However, in the paper by Eremets \textit{et al.}
\cite{Eremets} there is a residual resistance of 8 $\Omega$ in the silane
sample evident from Fig. 2b of Ref. \cite{Eremets}, while the Meissner
effect has not been measured. We argue that the superconductivity of the
"metallic hcp phase of silane" has not been shown convincingly. Moreover,
no plausible explanation has been given in Ref. \cite{Eremets} for
the pressure dependence of T$_{c}$ which is quite unusual for such
a simple hcp structure with low compressibility. Here however, we
offer our explanation of the experimental observations. Although the
superconductivity of PtH has not been measured in a separate experiment
and it is not known if it is superconducting, it is plausible that
the reported superconductivity \cite{Eremets} could originate from
the PtH formed in the diamond anvil cell. The quasi-four electrode
scheme used in the experiment of Ref. \cite{Eremets} ``involves
contribution of resistance of small piece of platinum foil", and this
Pt foil after forming PtH would show a drop in resistivity observed in
Ref. \cite{Eremets} if PtH were superconducting. Also, the formation of
PtH means a 15\% volume expansion of Pt (estimated for the pressure
of 113 GPa), and this would lead to a formation of a uniform PtH phase between
the Pt electrodes taking into account a small separation between electrodes
due to a very small size of the sample chamber in the experimental set-up
of Eremets et al (see Fig 2a of Ref. \cite{Eremets}). The peculiar
pressure dependence of the T$_{c}$ (the rise and the subsequent drop) can be
connected with the change of the hydrogen content in the hydride under
pressure (see discussion above). Pt-group metal hydrides (for example
Pd-H) are known to superconduct showing strong dependence of T$_{c}$
on the hydrogen content \cite{Stritzke}. In any case, the idea that PtH
might be superconducting is very interesting in itself and certainly
deserves further experimental attention.

In our own experiments on silane, we observe the SnBr$_{4}$ phase
(Ref. \cite{Degtyareva}) up to 50 GPa at room temperature. Above
this pressure, we detect a decomposition of silane into pure Si and
H$_{2}$. This is documented by the observation of the diffraction
signal from hcp Si (Fig. 3b). The strongest peak (101) of the hcp Si
is observed with a d-spacing that agrees with the reported Si lattice
parameters and equation of state \cite{silicon}. Hydrogen, released
during the decomposition of silane, reacts with the Re gasket and
participates in the formation of Re hydride, as discussed below. Earlier
theoretical calculations \cite{Pickard} reported that silane is unstable
to decomposition into hydrogen and silicon at pressures up to approx. 50
GPa, which supports our observations. We note that the very complex 
and kineticaly driven behavior of silane between 50-75 
GPa includes partial decomposition, pressure induced amorphization and 
a subsequent re-crystallization. These and other results will be described 
in detail in a separate paper \cite{Christophe,note2}.

Above 50 GPa at room temperature, we obtain a diffraction pattern that
contains an hcp phase from rhenium metal (used as a gasket material in
our diamond anvil cell setup) with an atomic volume of 13.02(1) \AA$^{3}$
at 65 GPa and another hcp phase similar to Re but with a larger atomic
volume of 13.84(1) \AA$^{3}$ (Fig. 3a). The "expanded Re" phase can
be identified as a Re hydride if one takes into account the fact that
hydrogen is known to increase the volume of a metal by approx. 2.1 \AA$^{3}$
per hydrogen atom (as discussed above). The atomic volume difference
between the pure and "expanded" Re phases of 0.82 \AA$^{3}$ gives a
composition of ReH$_{0.39}$. This corresponds within error to the known
composition (ReH$_{0.38}$) of the saturated Re hydride obtained above
8.6 GPa in which Re atoms form an hcp structure with a c/a ratio of 1.58
\cite{AtouBadding, ReHstr}, very close to c/a=1.57 of the hcp phase of
"expanded" Re obtained in our data. We observe that the expansion of Re
gasket due to the hydride formation (volume increase of about 6.3\%)
happens inward the sample chamber, meaning that Re hydride forms inside
the sample chamber (inset in Fig. 3c). We note that this would make the 
reflectivity measurements of silane at these pressures very difficult.

The obtained Re hydride is found to be stable up to 108 GPa, the maximum
pressure reached in this study (Fig. 3c), in accord with previous
studies on Re hydride to 120 GPa \cite{besedin}. Our data show that the
compressibility of Re hydride is very similar to that of Re. This agrees
with previous studies to 20 GPa that showed an apparent incompressibility
of the interstitial hydrogen, such that the difference in volume between
rhenium metal and rhenium hydride remained nearly constant. The Pt hydride
PtH with hcp structure remains stable up to at least 192 GPa, as can
be deduced from the stability of the "metallic hcp phase of silane"
from Ref. \cite{Eremets} that we have identified here as PtH, showing
compressibility very similar to that of pure Pt. For comparison, the
platinum carbide PtC, that is also synthesized under pressure, is reported
to remain stable in its fcc structure up to at least 120 GPa \cite{PtC}.

In our experiments at low temperatures of 100 K, the decomposition of
silane is observed at higher pressure of 60 GPa in comparison with the
decomposition pressure of 50 GPa at room temperature. The decomposition
of silane and the formation of metal hydrides are strongly dependent on
temperature as well as on time, which points to a kinetic origin of the
process. We also note that we have not observed the claimed "metallic
hcp phase of silane" \cite{Eremets} in any of our experiments.

\section{Conclusions}

Present high-pressure experiments on silane demonstrate the importance
of taking into account possible decomposition and chemical reaction
of the sample and surrounding materials when analyzing the data from
high-pressure measurements. Silane (SiH$_{4}$) is shown to partially
decompose releasing hydrogen which reacts with the Re gasket that
contains the sample inside the diamond anvil cell, forming Re hydride. The
"metallic hcp phase of silane" from Ref. \cite{Eremets} is identified
here as PtH \cite{Hirao} that forms upon the decomposition of silane and
reaction of released hydrogen with platinum metal that is present in the
sample chamber. A formation of tungsten hydride is reported in the studies
of silane and hydrogen mixture under pressure using a tungsten gasket
\cite{silanehydrogen}. In view of these observations, the reported 
metallization
of silane at pressures 50-60 GPa \cite{Eremets,Chen} should be critically
reconsidered. Our present measurements suggest that silane might not
metallise until much higher pressures \cite{Christophe}, in agreement
with theoretical predictions that pressures as high as 220-250 GPa might
be required to obtain metallic silane \cite{Pickard, MartinezCanales}.

On the other hand, silane turns out to be useful for the high-pressure
synthesis of metal hydrides. At pressures above 50 GPa, silane decomposes
releasing pure hydrogen which readily reacts with the surrounding metals
forming metal hydrides. The fact that silane is pyrophoric at normal
conditions makes it difficult to load. However, silane has a certain
advantage in comparison to other hydrogen-containing materials used as
internal hydrogen sources \cite{Antonov,Fukai}, as its decomposition
on compression does not require heating. This property may prove to be
useful for measurements of superconductivity of metal hydrides synthesized
under pressure, where heating above room temperature is undesirable.

\section{Acknowledgements}

The authors are grateful to M.I. McMahon and M. Mezouar for the help with
data collection, V.F. Degtyareva and J. Loveday for valuable discussions,
and N. Hirao for communicating results prior to publication. This work
is supported by a research grant from the U.K. Engineering and Physical
Sciences Research Council and facilities made available by the European
Synchrotron Radiation Facility. O.D. acknowledges support from the
Royal Society.

\begin{figure}
\includegraphics[scale=0.75]{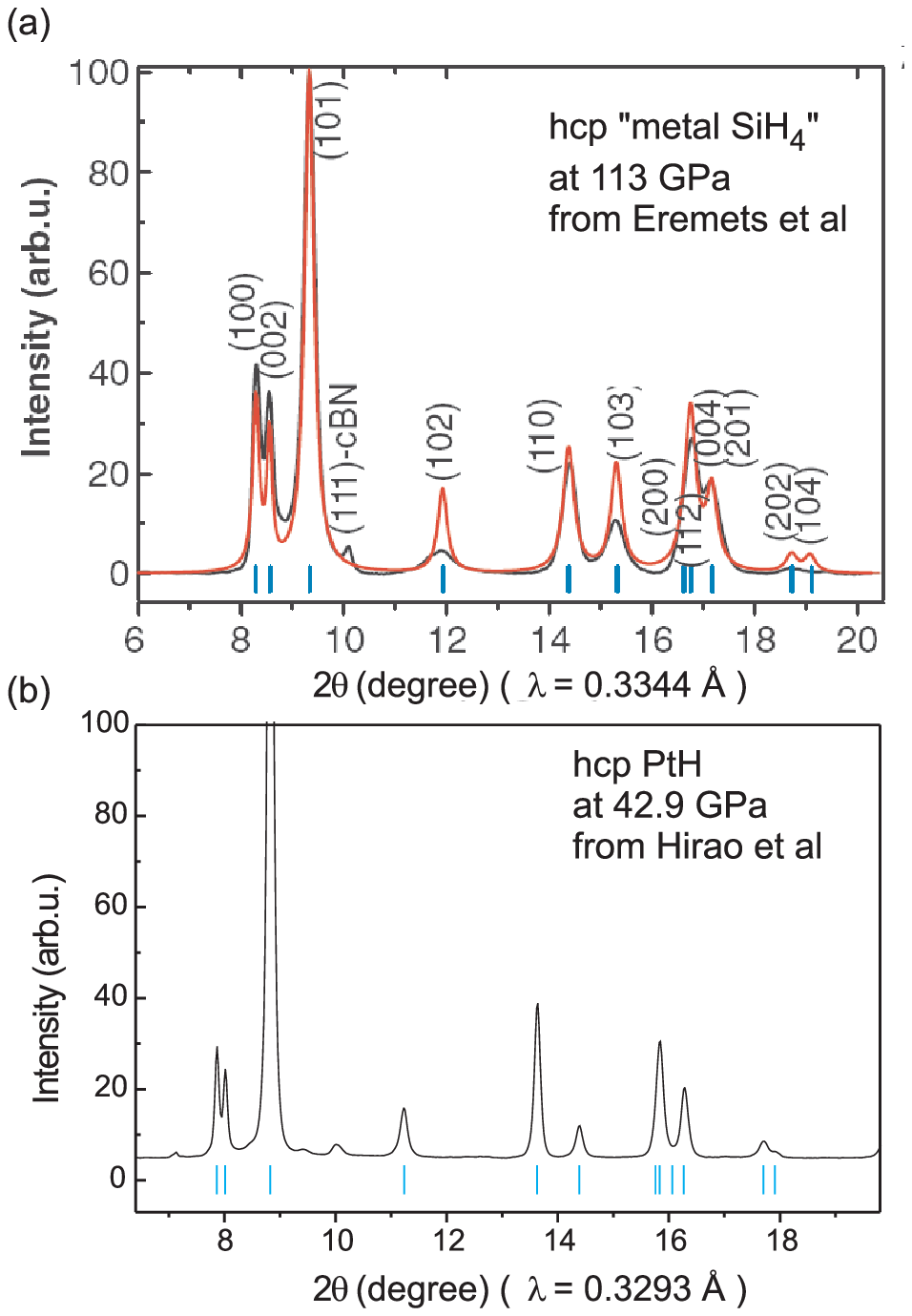}
\caption{Fig. 1.  Powder diffraction spectra of (a) the claimed "hcp 
phase of metallic silane" \cite{Eremets} and (b) hcp phase of PtH 
synthesised at high pressure \cite{Hirao}. The tick marks in (b) 
show peak positions calculated from the reported lattice parameters 
for PtH in its hcp phase: a = 2.730(1) \AA, and c = 4.288(1) \AA.}
\label{fig1}
\end{figure}

\newpage
\begin{figure}
\includegraphics[scale=1.5]{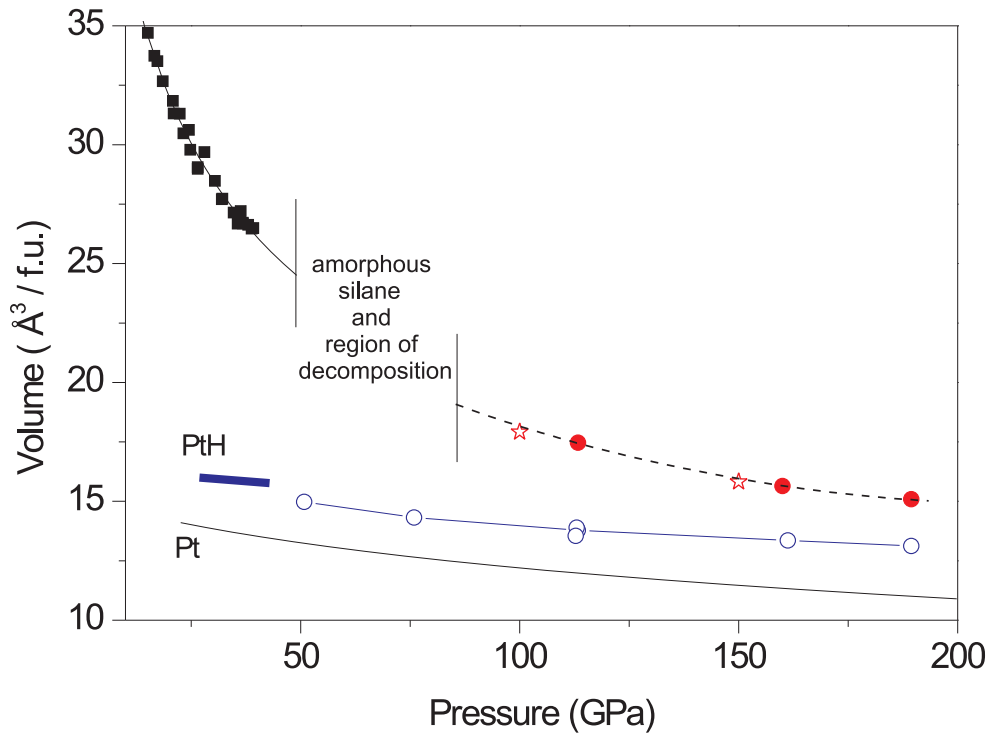}
\caption{Fig. 2.  Equation of state for silane shown in comparison with
PtH and Pt. Filled squares are from measurements on SnBr$_{4}$ phase
of silane from Ref. \cite{Degtyareva}, filled circles show measured
data on the I4$_{1}/a$ phase of silane from Ref. \cite{Eremets},
and stars show the predicted points for I${\bar4}2d$ and I4$_{1}/a$
phases \cite{Pickard}. The observed range of amorphous silane and region
of its partial decomposition are indicated by the vertical lines. The
thick line denoted as PtH shows the volume of the hcp phase of PtH from
Ref. \cite{Hirao}. The open circles are for the "hcp phase of silane"
from Ref. \cite{Eremets}. The solid line shows the equations of state
for pure Pt from Ref. \cite{Dewaele}.}
\label{fig1}
\end{figure}

\newpage
\begin{figure}
\includegraphics[scale=0.75]{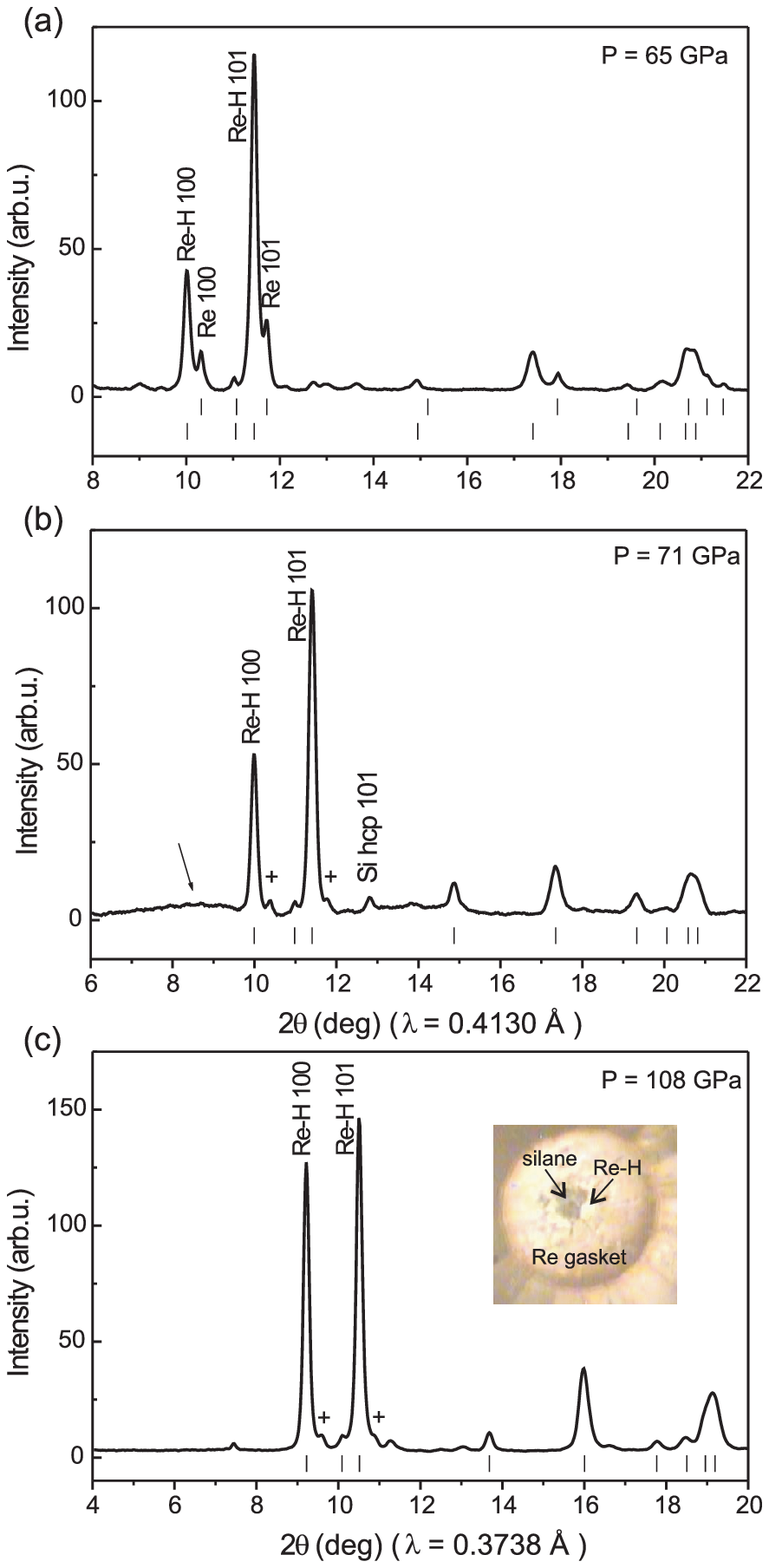}
\caption{Fig. 3.  X-ray diffraction spectra on Re hydride obtained in
the present work. (a) The spectrum at 65 GPa contains hcp phase of Re
(used as gasket material) shown with upper tick marks, and Re hydride
shown with the lower tick marks. The two most intense diffraction peaks
are marked with their hkl indices for both phases. (b) The spectrum at
71 GPa contains the Re hydride shown with tick marks, as well as the
hcp phase of Si (most intense peak is indicated). The arrow points to
a weak diffuse signal from amorphous silane with more information on
it given elsewhere \cite{Christophe}. "+" indicates the two strongest
peaks from a small amount of the hcp phase of Re. Other weak peaks in
(c) originate from the re-crystallized silane sample. Inset in (c) shows 
the micro-photograph of the (recrystallized) silane sample and Re 
hydride that has formed inside the sample chamber around the edges 
of the gasket hole. The dimension
of the figure is about 100 $\mu$m.}
\label{fig1}
\end{figure}
 

\begin{thebibliography}{00}
\bibitem{note} The spectra of the claimed "metallic phase of silane" 
and PtH (Fig. 1a,b)
are identical apart from the diffraction lines in Fig. 1a being shifted to
higher 2-$\theta$ angles in comparison to Fig. 1b. This is due to the
difference in pressures at which these spectra were measured. Taking into
account the compressibility of PtH which is close to that of Pt (Fig. 2),
the positions of diffraction lines of PtH in Fig. 1b would shift to higher 
2-$\theta$ angles and at pressure of approx. 113 GPa would overlap 
with those in Fig. 1a.
\bibitem{note2}The amorphous x-ray diffraction signal from the silane
sample observed in our work \cite{Christophe} is very weak, which
apparently was missed in the experiment of Ref. \cite{Eremets}, where
instead a stronger diffraction signal from crystalline PtH was measured.
We note that even a very small amount of PtH in the sample chamber
($\sim$5 $\mu$m) will give a much stronger x-ray diffraction than the
amorphous silane sample.
\bibitem{ReHstr}It has since been shown that the true structure of
Re hydride is a superstructure of hcp with a anti-CdI$_{2}$ structure
\cite{Kawamura}.
\end{thebibliography}
\end{document}